\documentclass[12pt]{article}
\usepackage[english]{babel}
\usepackage{amsmath}
\def\vk{\vert k\vert}
\def\di{D\!\!\!\!\slash\,}

\def\ha{\frac{1}{2}}
\def\tr{\,{\rm tr}\,}

\def\pr{\prime}
\def\mg{m_\gamma}
\def\be{\bar\eta}
\def\ba{\bar\alpha}
\def\t{\tilde}
\def\ga{\gamma}

\def\al{\alpha}

\def\eps{\epsilon}

\def\cd{{\cal D}}
\def\cp{{\cal P}}

\def\>{\rangle}
\def\<{\langle}
\def\pb{\bar{\psi}}
\def\Pb{\bar{\Psi}}

\def\di{D\!\!\!\!\slash\;}

\def\pa{\partial}
\def\pan{\par\noindent}

\def\qw{\qquad\hbox{where}\quad}
\def\qa{\qquad\hbox{and}\quad}

\begin{document}
\begin{titlepage}
\title {
\hfill{\small ETH-TH/91-15}\\[15mm]
Finite Temperature Schwinger Model}
\author{I. Sachs and A. Wipf\\[4mm]
\emph{Institut f\"ur Theoretische Physik}\\
\emph{Eidgen\"ossische Technische Hochschule}\\
\emph{H\"onggerberg, Z\"urich CH-8093, Switzerland}}

\date{\small 12 August 1991; arXiv-ed 10 May 2010}
\maketitle
\begin{abstract}
\noindent
The temperature dependence of the order parameter of the Schwinger model
is calculated in the euclidean functional integral approach. For that we
solve the model on a finite torus and let the spatial extension
tend to infinity at the end of the computations. The
induced actions, fermionic zero-modes, relevant Green functions
and Wilson loop correlators on the torus are derived. We find
the analytic form of the chiral condensate for any temperature and
in particular show that it behaves like $\<\Pb\Psi\>\sim -2 T\exp(-\pi
\sqrt{\pi}T/e)$ for temperatures large compared to the induced photon
mass.
\end{abstract}
\vskip5mm
\begin{center}
Keywords: Schwinger model; finite temperature; Euclidean path integral; fermionic zero modes; 
Wilson loop; condensate; effective action; field theory, torus; two-point function
\\[3mm]
Published as: Helvetica Physica Acta \textbf{65} (1992) 652-678
\end{center}

\hrule

\vskip4mm
{\small email (2010):  ivo.sachs@physik.lmu.de, wipf@tpi.uni-jena.de}\\[-2mm]
\hrule

\end{titlepage}
\tableofcontents

\section{Introduction}
\indent
The study of exactly soluble field theories has always
received a good deal of attention in the hope that
they might shed some light on more realistic theories.
One such model is the Schwinger model or quantum electrodynamics
for massless fermions in two space-time dimensions \cite{schwinger}.
It serves as an important tool in illustrating various (related)
field theoretical concepts such as mass generation, dynamical symmetry
breaking, charge shielding or fermion trapping \cite{coleman}.\par
More recently, the Schwinger model and its
extensions have been used as toy models for studying Baryon
number violating processes \cite{bocharev}. Since the semiclassical expansion
in the standard model breaks down at the interesting energies or
temperatures, it is important to compare the realistic calculations
to the corresponding ones for simpler models. As simple
soluble models which, for the problem at hand, have features identical
to the standard model, the non-linear sigma model \cite{mottola} and the
Schwinger model (coupled to a scalar field) \cite{bocharev} have been considered.
\par
The influence of an external gravitational field on the ultra-violet
and infra-red properties of the model has also been investigated \cite{gass}.
This problem is not completely solved up to now but the partial
results indicate that qualitatively all long-distance features of
the flat-space-time solution persist in curved backgrounds. \par
The zero-temperature Schwinger model has been solved some time
ago by using operator methods \cite{loewenstein} and more recently in the path integral
formulation \cite{nielsen}. Some properties of the model (e.g. the non-trivial
vacuum structure) are more transparent in the operator approach and
others (e.g. the role of the chiral anomaly) are better seen in
the path integral approach.\par
When studying the finite temperature $O(3)$-sigma
model coupled to fermions (as a relevant model for the $B$-number
violating processes in the electro-weak theory \cite{mottola}) we realized
that no satisfactory solution of the simpler finite-temperature Schwinger
model is available, at least in the path integral approach.
Therefore we decided to investigate the Schwinger model at finite temperature
and in particular determine the temperature dependence of the
chiral condensate by using functional integral methods. To study the
temperature effects we solve the Schwinger model on the two-dimensional
torus to avoid the infra-red divergences, which sometimes plague
the analysis of two-dimensional gauge theories.\par
Previous finite-volume solutions of $QED_2$ have clarified various
aspects of the model. For example the role of the nonintegrable
phase factor $\exp(ie\int A_\mu\,dx^\mu)$ has been understood by
studying the model on a cylinder, that is by considering
the zero-temperature model where the spatial dimension is given by a circle
\cite{manton}. Recently the Schwinger model has been solved in the path integral
approach on the 2-dimensional sphere and the role of the
fermionic zero modes has been emphazise \cite{jayewardena}. Also it has been
considered on a 2-dimensional disk assuming the most general (local)
self-adjoint boundary conditions for the fermions \cite{balog}.\par
However, these geometries
are not the proper ones to study finite temperature effects. For that
we must assume that the fields are (anti)-periodic in the time
direction with period $\beta=1/T$. In order to justify the manipulations
below (in particular the treatment of the Grassmannian integrals) we prefer
to work on a finite space-time volume such that the relevant wave-operators
possess discrete spectra. Thus we assume that the spatial extension
of space-time is finite as well and consider the Schwinger model on a torus
$[0,\beta]\times [0,L]$ with volume $V=\beta\cdot L$. We assume periodic
boundary conditions in the spatial direction. Only at the end of our
calculations do we let the spatial extension $L$ tend to infinity.\par
The only other solution of the Schwinger model on the torus we are
aware of, besides assuming periodic boundary conditions for Dirac-K\"{a}hler
fermions instead of finite temperature ones, concentrated mainly
on finding and discussing the fermionic zero-modes \cite{joos}.
Also, our trivialization of the $U(1)$-bundle over the torus is different
from the one in \cite{joos} but is very convenient for determining the explicit
spectrum of the Dirac operator.
\par
Our main results are the analytic form of the (zero-mode truncated)
effective actions and zero-modes in the various topological sectors,
the two-point Green function of $\Delta^2\!-\!\mg^2\Delta$
on the torus and as a consequence the explicit formulae (5.9), (5.10)
and (5.12) for the temperature dependence of the chiral condensate $\<\Pb\Psi\>$.
We find that $\<\Pb\Psi\>_T$  is equal to $-(\mg/2\pi)\cdot e^{\gamma}\;$
for $T=0$, in agreement with the zero-temperature solutions [7,9], and tends
to $-2T\exp(-\pi T/\mg)$ for temperature large compared to
the induced photon mass $\mg\!=\!e/\sqrt{\pi}\,$.
Our high temperature result differs by a factor $2$ from the only other
(formal) finite temperature path integral solution we are aware of \cite{kao}.
Also, the torus calculation nicely splits the chiral anomaly responsible
for the non-vanishing condensate, into its high- and low energy parts.
The first (coming from the gauge invariant ultraviolet regularisation) only affects
the numerical value of $\<\Pb\Psi\>$ , whereas the latter (coming from the chiral
asymetry of the zero-energy sector) is responsible for the non-zero $\<\Pb\Psi\>$.
This is the reason why ordinary perturbation theory, although leading to the correct
$UV$-contribution via the wellknown anomaly graph, fails to predict a non-vanishing condensate. 
At the end we obtain analytic expressions for (thermal) Wilson loop
correlators and in particular their temperature dependence as well as the gauge invariant fermionic 
two-point functions.

\section{The Schwinger Model}

\indent
In this section we recapitulate briefly the euklidean Schwinger model \cite{nielsen},
using functional methods. We emphasize the less known features involved
when considering the model on a torus instead on the plane. The Schwinger
model is defined by the action
\[
S[A,\Pb,\Psi]=\ha\int d^2x\,E^2-\int d^2x\,\Pb i\di\Psi,\eqno(1.1)
\]
where
\[
E=F_{01}=\pa_0 A_1-\pa_1 A_0\quad\hbox{and}\quad
\di=\gamma^\mu\big(\pa_\mu-ieA_\mu\big).\]
Note that in two space-time dimensions the electric charge
has the dimension of an inverse length, contrary to the
situation in four dimensions.
The model is superrenormalizable and requires no infinite renormalization
besides a trivial redefinition of the zero energy density. The generating
functional for the Green functions is
\[
Z[J,\eta,\be]={\cal C}^{-1}\int \cd A\cd \Pb\cd \Psi \;e^{-S[A,\Pb,\Psi]-S_{gf}[A]
+\int d^2x \big(A_\mu J^\mu+\be \Psi+\Pb\eta\big)},\eqno(1.2)
\]
where $S_{gf}[A]$ contains the terms due to gauge fixing
and $J^\mu$ resp. $(\be,\eta)$ are $c$-number and Grassman-valued
currents respectively. The normalization constant ${\cal C}$ is
chosen such that $Z[0,0,0]=1$.
First we evaluate the fermionic path integral
\[
Z[A;\eta,\be]=\int \cd \Pb\cd \Psi \;e^{\int \Pb i\di \Psi
+\int d^2x \big(\be \Psi+\Pb\eta\big)},\eqno(1.3)
\]
and treat the 'photon'-field as external field in this first step.\par
On the torus the Dirac operator may possess normalizable zero-modes,
$\di\psi=0$, and thus we must allow for such
fermionic zero modes in the path integral (1.3). Let us assume that
there are $\vk$ (our notation will become clear later on)
normalizable zero-modes $\psi_p$.
In addition there are an infinite number of excited modes
$\psi_q,\,q>\vk$ with eigenvalues $\lambda_q\neq 0$.
The number of zero-modes can be determined as follows:\par
First we note that since $\gamma_5$ anti-commutes
with the Dirac operator the excited modes come in pairs
\[
i\di\psi_q=\lambda_q\psi_q\Longrightarrow i\di\big(\gamma_5\psi_q\big)=
-\lambda_q\big(\gamma_5\psi_q\big),\eqno(1.4)\]
so that $\psi_q$ and $\gamma_5\psi_q$ have eigenvalues $\lambda_q$ and
$-\lambda_q$ and are orthogonal to each other.
It follows that the contribution of the excited states to the
(supersymmetric) partition function
\[
\tr\;\gamma_5\,e^{t\di^2}=\sum_1^{\vk}(\psi_p,\gamma_5\psi_p)+
\sum_{\vk+1}^\infty e^{-t\lambda_q^2}(\psi_q,\gamma_5\psi_q).
\]
vanishes. Thus the partition function is just the index $n_+\!-n_-\equiv k$,
where $n_+$ is the number of right handed (chirality $\gamma_5=1$) and
$n_-$ the number of left handed ($\gamma_5=-1$) zero modes.
On the other hand, inserting the small- $t$ expansion (see e.g. \cite{mckean})
\[
\<x\vert e^{t\di^2}\vert x\>\sim \frac{1}{4\pi t}\big(1+\gamma_5E t
+O(t^2)\big)\eqno(1.5)
\]
one immediately obtains the index theorem (see e.g. \cite{gilkey})
\[
k=\frac{1}{2\pi}\int d^2x \,E\equiv \frac{1}{2\pi}\Phi\eqno(1.6)
\]
which relates the index $k$ of the Dirac operator to the flux of the
electric field.
This formula already shows that there are zero-modes for non-vanishing fluxes
$\Phi$. Also note that on the torus the flux
is quantized in integer muliples of $2\pi$. This is really a consequence
of the single valuedness of the fermionic wave function (cocycle
condition).\pan
The pairing property (1.4) does not hold for the zero-modes.
Indeed, they can be chosen to have fixed chirality
since $\gamma_5$ commutes with the Dirac operator on the subspace spanned
by the zero modes.
Also (1.4) would not be valid for the excited modes if we would impose
chirality-violating boundary conditions as in \cite{balog}. But
(anti) periodic boundary conditions are compatible with the transformation
$\psi\to \gamma_5\psi$ and the identity (1.4) holds in the present situation.\pan
To determine $n_+$ and $n_-$ separately we decompose the gauge potential as
\[
A_\mu=\tilde A_\mu-\eps_{\mu\nu}\pa_\nu\phi\quad\hbox{where}
\quad \phi=\frac{1}{\Delta}\big(E-\frac{1}{V}\Phi\big)+\hbox{c}\Longrightarrow
E=\frac{\Phi}{V}+\Delta \phi,\eqno(1.7)
\]
that is into a global "instanton"-type potential $\tilde A$ with
constant field strength, $\tilde E=\Phi/V$ and a local
fluctuation $\delta A_\mu=-\eps_{\mu\nu}\pa_\nu\phi$ about the instanton.
The subtraction of the constant term in (1.7) is necessary since
on the torus the Laplacian has the constant zero-mode and
is invertible only on functions which integrate to zero. Using
$\gamma_\mu\gamma_5=-i\eps_{\mu\nu}\gamma_\nu$ it is now easy to see that
\[
\di=e^{\gamma_5\phi}\tilde\di e^{\gamma_5\phi}\eqno(1.8)
\]
which shows that the number of fermionic zero-modes is independent
of $\phi$ and hence is the same for $A$ and $\tilde A$. But for the
instanton potentials $\tilde A$
\[
\tilde\di^2=\tilde D^2+\gamma_5\,\frac{\Phi}{V}\eqno(1.9)
\]
and since $-\tilde\di^2$ is a non-negative operator, all zero modes
are either right- or left-handed for non-vanishing fluxes. Only in the
zero-instanton sector $\Phi\!=\!0$ can it happen that there are
both right- and left-handed zero modes. For example, for a vanishing
gauge potential there is one right-handed and one left-handed zero mode.
\pan What we have shown then is that for $\Phi\neq 0$
there are exactly $\vk=\vert \Phi\vert/2\pi$ zero modes. If the
flux is positive they are all right-handed, else they are all left-handed.\par
After having counted the number of zero modes we proceed
by expanding the 'electron'-field in an adapted orthonormal eigen-base as
\[
\Psi(x)=\sum_1^{\vk}\al_p\psi_p(x)+\sum_{\vk+1}^\infty \beta_q\psi_q(x)\]
and similarly $\Pb$, so that
\[
(\bar\eta,\Psi)=\sum (\bar\eta,\psi_p)\al_p+\sum (\bar\eta,\psi_q)\beta_q
\qquad (\Pb,\eta)=\sum \bar\al_p(\pb_p,\eta)+\sum\bar\beta_q(\pb_q,\eta)\]
split into a zero-mode part and an excited part.
Inserting this decomposition into (1.3) and using $\cd\Pb\cd\Psi=
\cd\ba\cd\al\cd\bar\beta\cd\beta$ the Grassmannian
integral over the $\al$'s can easily be done since the action does not depend
on them. This way one finds for the zero-mode contribution to (1.3)
\[
\int \cd\ba\cd\al\,\prod_1^{\vk} e^{(\bar\eta,\psi_p)\al_p}\,e^{\bar\al_p
(\pb_p,\eta)}=\prod_1^{\vk}(\bar\eta,\psi_p)(\pb_p,\eta).\eqno(1.10a)
\]
The remaining $\beta$-integration
is performed by shifting the $\beta$'s (and similarly the $\bar\beta$'s)
according to
\[
\beta_q\longrightarrow \beta_q-\frac{1}{ \lambda_q}(\pb_q,\eta).\]
After this shift the $\beta$ integration yields
\[
\int \cd\bar\beta\cd\beta \prod_{\vk+1}^\infty e^{\lambda_q\bar\beta_q
\beta_q+(\be,\psi_q)\beta_q+\bar\beta_q (\pb_q,\eta)}
=e^{-\int\bar\eta(x)G_e(A;x,y)\eta(y)}\cdot{\det}^\pr (i\di),\eqno(1.10b)
\]
where $\det^\pr$ is the determinant with the zero-eigenvalues
omitted and
\[
G_e(A;x,y)=\sum_{\vk+1}^\infty \frac{\psi_q(x)\psi_q^\dagger(y)}{ \lambda_q}.
\eqno(1.11a)
\]
is the Green function on the space orthogonal to the
zero modes. Clearly this function obeys the differential equation
\[
i\di G_e(A;x,y)=\delta(x\!-\!y)-P(x,y),\qw P(x,y)=\sum_1^{\vk}
\psi_p(x)\psi^\dagger_p(y)\eqno(1.11b)
\]
is the projection kernel on the zero-mode subspace.
Due to the pairing property of the excited modes
the 'excited' green function anti-commutes with $\gamma_5$:
\[
\gamma_5 G_e(A;x,y)\gamma_5=-G_e(A;x,y).\eqno(1.11c)
\]
Inserting now (1.10) into the path integral for the (fermionic) partition
function (1.3) we end up with
\[
Z[A;\bar\eta,\eta]=\prod_1^{\vk}\,(\bar\eta,\psi_p)(\pb_p,\eta)\;
\;e^{-\int\bar\eta(x) G_e(A;x,y)\eta(y)}\;{\det}^\pr(i\di).\eqno(1.12)
\]
Thus to determine $Z(A;\be ,\eta)$ for a fixed potential we need to compute
the $\vk$ zero-modes of the Dirac operator,
the determinant of $i\di$ with the zero-modes omitted and the 'excited'
Green function.
The (unnormalized) fermionic $2n$-point functions for a given $A$-field
are now obtained by differentiation with respect to the external currents
\[
\int \cd\Pb\cd\Psi\;\Psi_{\al_1} (x_1)\Pb_{\beta_1}
(y_1)...\Psi_{\al_n} (x_n)\Pb_{\beta_n} (y_n)\,e^{\int \Pb i\di \Psi}\]
\[ \quad =
\frac{\delta^{2n}}{ \delta\eta^{\beta_n}(y_n)\delta\be^{\al_n}(x_n)
...\delta\eta^{\beta_1}(y_1)\delta\be^{\al_1} (x_1)}Z[A;\be,\eta]
\vert_{\be =\eta =0}\eqno(1.13)
\]
Using (1.12) in this formula we can immediately read off that
\begin{enumerate}
\item The fermionic partition function
\[
Z[A;0,0]={\det}(i\di)\eqno(1.14)
\]
which enters in the normalization constant ${\cal C}$ in (1.2)
is only non-zero if the Dirac operator possesses no zero modes that is for
gauge field with vanishing flux.
\item The $2$-point functions are non-zero only if $A$ admits either
no or one zero-mode, that is for $\Phi=0$ or $\Phi=\pm 2\pi$ and then
\[
\int \cd\Pb\cd\Psi \;e^{\int \Pb i\di\Psi}\,\Pb_\al (x)\Psi_\beta(y)=
\begin{cases}
-G_{\al\beta}(x,y){\det}(i\di) & \text{$\Phi=0$}\\
-\psi^\dagger_\al(x)\psi_\beta(y){\det}^\pr(i\di) &  \text{$\vert\Phi\vert=2\pi$.}
\end{cases}
\eqno(1.15)
\]
From (1.11c) it follows that the expectation values of
$\Pb P_\pm\Psi$, where $P_\pm=\ha (1\pm\gamma_5)$ are the chiral
projections, are non-vanishing only for $\Phi=\pm 2\pi$ in which
cases the Dirac operator has one zero mode $\psi$ of chirality
$\pm 1$, and then
\[
\int \cd\Pb\cd\Psi \;e^{\int \Pb i\di\Psi}\,\Pb (x)P_\pm \Psi(x)=
-\tr\big(\psi^\dagger (x)P_\pm\psi(x)\big)\,{\det}^\pr (i\di).\eqno(1.16)
\]
The higher $2n$-point functions are obtained similarly. For example,
for the $4$-point functions only gauge potentials with no, one or two
zero modes contribute. In particular for the expectation value
of the operator with (chiral) charge zero
\[
O(\Psi)=(\Pb P_+\Psi)(\Pb P_-\Psi),\qquad O(e^{\al\gamma_5}\Psi)=O(\Psi)
\eqno(1.17)
\]
only the trivial sector with no zero-modes yields a non-vanishing result.
This explains why earlier solutions of the Schwinger model on the plane
(in which the non-trivial sectors have been neglected) yield
the correct result for this expectation value and thus
(via clustering) the correct result for the chiral condensate.
\end{enumerate}
Note that the excited fermionic Green function does not appear in the
expectation values (1.16), and this simplifies the computation
of the chiral condensate considerably. More generally, the expectation
values of the charge $\pm n$ operators $(\bar\Psi P_\pm\Psi)^n$
are non-zero only for potentials with fluxes $\Phi=\pm 2\pi n$,
and then only the zero modes and effective actions, but not the
excited Green functions occur in the expectation values.

\section{Effective Actions - Local Part}
In this section we compute the $\phi$-dependence of the effective actions
$\det^\pr i\di$ for the general gauge potentials (1.7) by integrating
the chiral anomaly. Since there are some subtleties when
zero-modes are present we derive the local $\phi$-dependent part of the
effective actions, although some of our arguments are not new \cite{blau}.\pan
To determine the $\phi$-dependence of the effective actions
we consider the one-parametric family of Dirac-operators
$$
\di_\al=e^{\gamma_5\phi\al}\tilde\di e^{\gamma_5\phi\al}
\Longrightarrow \dot{\di}_\al=\{\gamma_5\phi,\di_\al\}
\qquad (\;\dot{}=\frac{d}{d\al})\eqno(2.1)
$$
which interpolates between $\tilde \di$ and $\di$, and calculate
the variation of the zeta-function regularized determinants \cite{seiler}
$$
\log{\det}^\pr (i\di_\al)=\ha \log{\det}^\pr(-\di^2_\al)=
-\ha \frac{d}{ds}\zeta_{\di^2}(s)\vert_{s=0}.\eqno(2.2a)
$$
The zeta-function is defined as as the analytic continuation of
$$
\zeta_{\di^2}(s)=\sum_{\vk+1}^\infty\mu_q^{-s},\qquad \Re(s)>1,\eqno(2.2b)
$$
where $\mu_q\!=\!\lambda_q^2$ are the $\al$-dependent excited eigenvalues
of the squared Dirac operator, to the whole complex $s$-plane.
Using (2.1) in the Feynman-Hellman formula
$$
\dot{\lambda}_q=(\psi_q,i\dot{\di}\psi_q)=2\lambda_q
(\psi_q,\gamma_5\phi \psi_q)
$$
the variation of the zeta-function can be written as
$$
\frac{d}{d\al}\zeta_{\di^2_\al}(s)=-s\sum_{\vk+1}^\infty \mu_q^{-s-1}
\dot{\mu}_q=-4s\sum\mu_q^{-s}(\psi_q,\gamma_5\phi\psi_q).
$$
This can be further rewritten as a Mellin transform
$$
\frac{d}{d\al}\zeta_{\di^2_\al} (s)=-\frac{4s}{\Gamma(s)}
\int dt\, t^{s-1}\sum_{\vk+1}^\infty e^{-t\mu_q}(\psi_q,\gamma_5\phi\psi_q)
=-\frac{4s}{\Gamma(s)}\int dt\,t^{s-1}\tr^\pr\Big(e^{t\di^2_\al}\gamma_5
\phi\Big),
$$
where the trace is only to be taken over the excited states. Inserting
the asymptotic expansion (1.5) from which we must subtract the
projection density $P_\al (x,x)=P_\al(x)$ (see (1.11b) onto the zero modes
we find
$$
\frac{d}{d\al}\frac{d}{ds}\zeta_{\di^2}(s)\vert_{s=0}=
-\frac{2\al}{ \pi}\int d^2x\,E\phi+4\int \tr\big(P_\al(x)\gamma_5
\phi(x)\big).\eqno(2.3)
$$
To integrate over $\al$ we observe that the zero-modes
of $\di_\al$ and $\tilde\di$ are related as
$$
\psi_p^{(\al)}=e^{\mp\al \phi}\tilde\psi_p,\qquad p=1,\dots,\vk \eqno(2.4)
$$
with the negative sign in the exponent for right-handed and
the positive sign for left-handed zero-modes. These modes are in
general not orthonormal, and the normmatrix
$$
{\cal N}_{pr}(\al)=\big(\psi_p^{(\al)},\psi_r^{(\al)}\big)\eqno(2.5)
$$
is not the identity so that the projection density reads
$$
P_\al(x)=\sum_{pr}\psi_p^{(\al)}(x){\cal N}^{-1}_{pr}(\al)\,(\psi_r^{(\al)})
^\dagger(x).
$$
Now it follows from (2.4) and (2.5) that
$$
\frac{d}{d\al}\log\det\big({\cal N}(\al)\big)=-2\int d^2x \tr\big(P_\al
\gamma_5\phi\big)\eqno(2.6)
$$
and this formula can be used to integrate the anomaly equation (2.3)
from $\al=0$ to $\al=1$. Together with (2.2) we end up with
$$
{\det}^\pr (i\di)=\det \frac{\cal N}{\tilde{\cal N}}\,{\det}^\pr(i\tilde\di)
\;\exp\big(\frac{e^2}{ 2\pi}\int E\phi\big)\eqno(2.7)
$$
which is the (almost) factorization of the determinants into a global,
$\tilde A$-dependent, and local, $\phi$-dependent, part. The factorization
is not complete since $\det{\cal N}=\det{\cal N}(1)$ still contains a coupling
between the instanton potential $\tilde A$ and the local fluctuation
$\delta A$ via (2.4). Since $E$ in (1.7) depends on both $\Phi$
and $\phi$, there is also an apparent coupling in the last factor
in (2.7). However we shall see later that not all functions $\phi$
are permitted and that for the allowed ones this factor does
not depend on $\Phi$. Also note that the local part does not
depend on the geometry of the torus. In particular the induced
photon mass $\mg=e/\sqrt{\pi}$ is the same as on the infinite plane.

\section{Effective Actions - Global Part}
With the factorizations (2.4) and (2.7) of the zero modes and effective
actions into local and global factors the problem of finding the
$2$-point functions (1.16) (or $2n$-point functions of charge $\pm n$
operators) reduces to the
problem of computing the zero modes and effective actions for the instanton
potentials $\tilde A$.
For that we need to know the explicit form of these potentials.\pan
An instanton potential can always be decomposed as
$$
\tilde A_0=-\frac{\Phi}{ V}x^1+\frac{2\pi}{ \beta}h_0+\pa_0\lambda\quad\hbox{and}
\quad\tilde A_1=\frac{2\pi}{ L}h_1+\pa_1\lambda \eqno(3.1)
$$
with constant $h_\mu$. The least trivial term proportional to
$\Phi$ yields the constant field strength and corresponds to a particular
trivialization of the $U(1)$-bundle over the torus. In other words, the
gauge potentials at $(x^0,x^1)$ and $(x^0,x^1\!+L)$ are necessarily related
by a non-trivial gauge transformation
$$
A_{\mu}(x^0,x^1\!+\!L)-A_{\mu}(x^0,x^1)=\pa_\mu\al,\qw \al=-\frac{\Phi}{\beta}x^0.
\eqno(3.2a)
$$
Accordingly the fermionic wave functions transform as
$$
\psi (x^0,x^1\!+\!L)=e^{i\al}\psi(x^0,x^1).\eqno(3.2b)
$$
These boundary conditions must be supplemented by the
finite temperature boundary conditions in the $x^0$-direction
$$
A_{\mu}(x^0\!+\!\beta,x^1)=A_{\mu}(x^0,x^1)\qa
\psi (x^0\!+\!\beta,x^1)=-\psi(x^0,x^1).\eqno(3.2c)
$$
The constant terms in (3.1) are the harmonic pieces of the gauge potential.
Even in the trivial sector $\Phi=0$ they yield the (dynamical)
nonintegrable phase factors
$$
e^{i\int A_\mu dx^\mu}=\exp\big[2\pi i(h_0 n_0+ h_1 n_1)\big]\eqno(3.3)
$$
for loops which wind $n_0$-times around the torus in the $x^0$ direction
and $n_1$-times in the $x^1$-direction.\pan
The last term in (3.1) is a pure gauge term and will drop
out at the end of the calculations.\pan
Now we consider the sector with $\Phi=0$ and those with $\Phi\neq 0$ in
turn:\pan
\subsection{Trivial sector $\Phi=0$}
In this sector $-\tilde \di^2=(\nabla-h)^2 Id\;$ is simple
and possesses the double degenerate eigenvalues
(of course, the spectrum is not affected by the gauge term in (3.1))
$$
\mu_m=\big(\frac{2\pi}{\beta}\big)^2(m_0-a_0)^2+\big(\frac{2\pi}{ L}\big)^2
(m_1-a_1)^2\qw (a_0,a_1)=(\ha+h_0,h_1).\eqno(3.4)
$$
The corresponding zeta-functions $\zeta(s)=\sum \mu^{-s}_m$
are the generalized Epstein functions \cite{epstein}.
Applying the Poisson-resummation formula the derivative at $s\!=\!0$ can
be expressed in terms of Jacobi theta-functions as \cite{blau2}
$$
\frac{d}{ds}\zeta(s)\vert_{s=0}=-2\log\Big\vert \frac{1}{ \eta(i\tau)}
\Theta\Big[{\ha+a_0\atop \ha-a_1}\Big](0,i\tau)\Big\vert,\qw\tau=L/\beta,
\eqno(3.5a)
$$
is the ratio of the two circumferences of the torus and
$$
\eta(i\tau)=q^{1/24}\prod_{n>0}(1-q^n);\qquad q=e^{-2\pi \tau}\eqno(3.5b)
$$
Dedekind's eta-function. We have adopted the conventions
in \cite{mumford} for the theta-functions:
$$
\Theta\Big[{a\atop b}\Big](z,i\tau)=\sum_{Z}e^{-\pi\tau(n+a)^2+2\pi i(n+a)
(z+b)}.\eqno(3.5c)
$$
Taking the degeneracy of the eigenvalues
into account, the zeta-function regularized determinants (2.2a)
in the sector with no zero modes are
$$
\det(i\tilde\di)=\Big\vert \frac{1}{ \eta(i\tau)}
\Theta\Big[{\ha+a_0\atop \ha-a_1}\Big](0,i\tau)\Big\vert^2.\eqno(3.6)
$$
\subsection{The non-trivial sectors}
We have seen that the excited eigenmodes of the Dirac operator come
in pairs with opposite eigenvalues. Since $\gamma_5$ commutes with
the squared Dirac operator the chiral projections $P_\pm\psi_q$
of these modes are eigenmodes of $-\di^2$. Thus the excited
eigenmodes of the squared Dirac operator come in pairs as well and two
partners have
the same energies $\mu_q=\lambda_q^2$ but opposite chiralities. We have
also shown that there are exactly $\vk=\vert\Phi\vert/2\pi$ chiral zero modes
of the Dirac operator and hence of $-\di^2$. Since $\tilde\di^2$ in (1.19)
differs only by the constant $2\Phi/V$ in the two
chiral sectors, the $\vk$ zero modes with chirality, say $+1$, are
at the same time excited modes with energies $2\Phi/V$ and
chirality $-1$. Due to pairing there are also $\vk$ excited modes
with chirality $+1$, and so on. Thus $-\tilde\di^2$ possesses
the following spectrum
\[
\mu_n=\begin{cases}
0&\text{degeneracy $=\vert\Phi\vert/2\pi$}\\
2n\vert\Phi\vert/V & \text{degeneracy $=\vert\Phi\vert/\pi$}
\end{cases}
\eqno(3.7)\]
Note that, contrary to the situation in the zero-instanton sector,
the spectrum does not depend on the harmonic part $h$ of
the gauge potential $\tilde A$.\pan
The zeta-function is now
proportional to the ordinary Riemann zeta-function and one obtains
$$
{\det}^\pr (i\tilde\di)=\Big(\frac{\pi V}{\vert\Phi\vert}\Big)^{\vert\Phi
\vert/4\pi}\eqno(3.8)
$$
for the determinant of the non-zero eigenvalues.\par
Inserting (3.6) and (3.8) into (2.7) we end up with the following
formulae for the determinants in the sectors with $\vk$
zero modes:
\[
\det(i\di)=\Big\vert \frac{1}{\eta(i\tau)}
\Theta\Big\vert^2 \;\exp\Big(\frac{\mg^2}{ 2}\int \phi\Delta \phi\Big)
\qquad\qquad\qquad\quad\qquad \;\;k=0\]
\[{\det}^\pr(i\di)=\big(\frac{V}{ 2\vk}\big)^{\vk/2}
\det\frac{{\cal N}}{ \tilde{\cal N}}
\; \exp\Big(\frac{\mg^2}{ 2}\int \phi\Delta\phi + \frac{e k}{ V}
\int \phi\Big) \qquad k\neq 0,\]
where $\Theta$ is the theta-function in (3.6) and we have expressed
$E$ in terms of $\phi$ and $\Phi$ (see (1.7)) and made the dependence
on the dimensionful electric charge explicit (recall that $\mg=e/\sqrt{\pi}$).

\section{Computing the Zero-Modes}
According to (2.4) we only need to evaluate the zero-modes
of $-\tilde\di^2$. Also we shall only consider expectation values
of gauge invariant operators and thus may set the gauge part
in (3.1) to zero. Since the gauge potentials (3.1) do not depend
on time the eigenmodes should be proportional to $\exp(ip_0x^0)$.
The finite temperature boundary conditions (3.2c) then require that the
momentum is quantized as $p_0=(2p\!-\!1)\pi/\beta$ with integer $p$.
If we further eliminate the $h_1$-part of the instanton potential
we are lead to the following ansatz
$$
\tilde\chi_p=e^{i(2p-1)\pi x^0/\beta}\,e^{i 2\pi h_1x^1/L}\;\xi_p(x^1)
$$
for the zero modes. Inserting this ansatz into the zero mode equation
$\tilde\di^2\tilde\chi_p=0$ yields
$$
\big(\frac{d^2}{ dy^2}-\frac{\Phi^2}{ V^2}y^2+\frac{\vert\Phi\vert}{ V}\big)
\xi_p =0,\qw y=x^1+\frac{L}{ k}(p-a_0)
$$
which shows that $\xi_p$ is the ground state wave function of a
harmonic oscillator and thus
$$
\xi_p=\exp\Big[-\frac{\vert\Phi\vert }{ {2V}}\big\{x^1+\frac{L}{ k}(p\!-\!a_0)\big\}^2
\Big],\eqno(4.1)
$$
where we have used the index theorem (1.6).
These functions do not obey the boundary condition (3.2b) but the correct
eigenmodes can be constructed as superpositions of them. For that one
notes that
$$
\tilde\chi_p(x^0,x^1\!+\!L)=e^{-i\Phi x^0/\beta}\,e^{2i\pi h_1}\;
\tilde\chi_{p+k}(x^0,x^1)\eqno(4.2)
$$
so that the sums
$$
\tilde\psi_p =
\big(\frac{2\vk}{ \beta^2 V}\big)^{1/4}
\sum_Z\,e^{2i\pi n h_1}\tilde\chi_{p+ n k},\qquad
p=1,\dots,\vk\eqno(4.3)
$$
obey the boundary conditions and thus are the $\vk$ required zero-modes
of the Dirac operator. The overall
factor normalizes these functions to one (the norm can be computed by
using 4.2)). Also, modes with different $p$ are orthogonal to
each other, so that the system (4.3) forms an orthonormal basis of
the zero-mode subspace. These functions
can be written in terms of Jacobi theta functions as (we drop a constant
phase)
$$
\tilde\psi_p=
\big(\frac{2\vk}{ \beta^2 V}\big)^{1/4}e^{2\pi i[h_0 x^0/\beta- k x^0x^1/V]}\;\;
\Theta\Big[{x^1/L+(p\!-\!a_0)/k\atop k x^0/\beta+a_1}\Big](0,i\vk\tau),
\eqno(4.4)
$$
where $\tau$ has been introduced in (3.5a). Since these modes
are orthonormal the determinant of the norm matrix $\tilde{\cal N}$ in (3.9)
is just one. The unnormalized zero modes which enter
the normmatrix ${\cal N}$ can now be obtained from the zero modes
(4.4) by the transformation (2.4) with $\al\!=\!1$.
These completes our discussion of the zero-mode sector. We have
computed all $\vk$ zero modes of $\di$ explicitly in terms of
Jacobi-theta functions.\pan
For the two-point functions of interest (1.16) only the
sector with one zero mode contributes, and since then the normalized
zero mode is just $\psi/\sqrt{{\cal N}}$, the normalization
constant cancels with ${\cal N}$ in (3.9) so that
$$
\int \cd\Pb\cd\Psi \;e^{\int \Pb i\di\Psi}\,\Pb (x)P_\pm\Psi(x)=
-\frac{1}{ \beta}\,\vert \Theta_\pm\vert^2\,e^{\mp 2\phi}
\; \exp\Big(\frac{\mg^2}{ 2}\int \phi\Delta\phi +\frac{e k}{ V}
\int \phi\Big),\eqno(4.5a)
$$
where $\Theta_\pm$ are the theta-functions in (4.4) with $p\!=\!0$ and
$k=\pm 1$:
$$
\Theta_\pm=\Theta\Big[{x^1/L\mp a_0)\atop x^0/\beta\pm a_1}\Big](0,i\tau)
\eqno(4.5b)
$$
(we used that the modulus of this function is unchanged if the sign
of the "lower parameter" is changed) and the $\pm$ sign refers to the
chirality of the zero modes.\pan
Our trivialization of the $U(1)$ bundle which leads to the form
(3.1) of the instanton potential differs from the one chosen in
\cite{joos} and so do our zero modes. But one can show \cite{sachs} that the modes in \cite{joos}
are obtained from our modes (4.4) by the corresponding change of
trivialization, as required.

\subsection{Calculating the bosonic path integral}
After having solved the fermionic integration we are now left with
the functional integrals over the $A$-field. In particular we
shall evaluate the two point functions
$$
\<\bar\Psi P_\pm\Psi\>=\frac{\int \cd A\,e^{-\ha\int E^2}\int \cd\Pb\cd\Psi
\;e^{\int \Pb i\di\Psi}\,\Pb (x)P_\pm\Psi(x)}{
\int \cd A\,e^{-\ha\int E^2}\int \cd\Pb\cd\Psi
\;e^{\int \Pb i\di\Psi}},\eqno(5.1)
$$
where $E$ is given in terms of $\phi$ and $\Phi$ in (1.7). Clearly,
since the integrands are expressed in terms of $\{\phi,h_\mu,\Phi\}$ it is
natural to change the integration variables from $A$ to these
variables.\pan
First one notes that there is a one to one correspondence between
$E$ and $\{\Delta\phi,\Phi\}$. This becomes a one to one correspondence
to $\{\phi,\Phi\}$ if we demand that $\phi$ is orthogonal to the kernel
of $\Delta$, that is if it integrates to zero. Second, for given
$\{\phi,\Phi\}$ the nonintegrable phase factors (3.3) are in one to one
correspondence to the phases $\exp(2i\pi h_\mu)$, and thus to the
$h_\mu$ modulo $1$.
Furthermore, from $\pa A=\Delta \lambda$ we see that there
is also a one to one correspondence between the divergence
of $A$ and functions $\lambda$ which integrate to zero. To summarize:
we have shown that the transformation
$$
A_\mu\longrightarrow \{\phi,\lambda,h_\mu,\Phi\}\qw \int \phi=\int\lambda=0
\qa 0\leq h_\mu<1,
\eqno(5.2)
$$
defined in (1.7) and (3.1), is one to one.\pan
Next we need to calculate the Jacobian of this transformation. This is
conveniently done by expanding all fields in eigenmodes of $-\Delta$,
e.g.
$$
\phi(x)=\frac{1}{ \sqrt{V}}\sum_{m\neq 0}\phi_m\,e^{2i\pi(m_0x^0/\beta+m_1x^1/L)},
\eqno(5.3)
$$
where $m=(m_0,m_1)=0$ is excluded because of the constraint on $\phi$
in (5.2). In terms of the coefficients the transformations (1.7) and
(3.1) read for $\Phi=0$
\[\hskip6mm
\begin{pmatrix}A_{0,m}\\ A_{1,m}\end{pmatrix}=\frac{2i\pi}{ L}
\begin{pmatrix}-m_1&\tau m_0\\
\tau m_0&m_1\end{pmatrix}\begin{pmatrix}\phi_m\cr\lambda_m\cr\end{pmatrix}\quad (m\neq 0)\]
\[\begin{pmatrix}A_{0,0}\\ A_{1,0}\end{pmatrix}=2\pi\begin{pmatrix}\sqrt{\tau}& 0\\
0&1/\sqrt{\tau}\end{pmatrix}\begin{pmatrix}h_0\\ h_1\end{pmatrix},\hskip14mm\]
and the Jacobian of this transformation is just $J=(2\pi)^2
{\det}^\pr(-\Delta)$ and thus independent of the dynamical fields.
In the nontrivial sectors labelled by $k$ we can
write $A_\mu=-\Phi/V\cdot x^1\delta_{\mu 0}+\delta A_\mu$ and
the above transformation applies then to $\delta A$, so that
$$
\int \cd A_\mu=\sum_k\cd \delta A_\mu=J\sum_k\int\limits_0^1 dh_0 dh_1
\int \prod_{m\neq 0}d\phi_m d\lambda_m.\eqno(5.4)
$$
Since in expectation values the same Jacobian appears in the numerator
and denominator we may neglect $J$ in what follows. Also, gauge invariant
operators (like in (5.1)) are independent of the gauge function $\lambda$ and
the integration over the $\lambda_m$ drops in expectation values as well
(in our decomposition $\lambda=0$ corresponds to the ghost free Lorentz
gauge).
Inserting (3.9) and (4.5a) into (5.1) we find
$$
\<\bar\Psi P_\pm\Psi\>=\frac{-1}{\beta}e^{-2\pi^2/e^2V}\;
\frac{\int d^2h\vert \Theta_\pm\vert^2\int \cd \phi \;e^{-\Gamma[\phi]\mp 2e\phi(x)}
}{ \int d^2h\vert \frac{1}{\eta}\Theta\vert^2\int \cd\phi \,e^{
-\Gamma[\phi]}}
\eqno(5.5a)
$$
where
$$
\Gamma[\phi]=\ha\int\phi(\Delta^2-\mg^2\Delta)\phi\eqno(5.5b)
$$
and one integrates over $\phi$ which integrate to zero.
$\Theta_\pm$ and $\Theta$ are the theta functions in (4.5b) and (3.6),
repectively. Note that the last term in the exponent in (4.5a) is absent
for the allowed $\phi$'s.\pan
Finally, using
$$
\int d^2h\vert\Theta_\pm\vert^2=
\int d^2h\vert\Theta\vert^2=\sqrt{\frac{1}{2\tau}}\eqno(5.5c)
$$
and performing the Gaussian functional integral yields
$$
\<\bar\Psi P_\pm\Psi\>=
-\frac{\vert\eta(i\tau)\vert^2}{ \beta}\,e^{-2\pi^2/e^2 V}\,e^{2e^2K(x,x)},
\quad\hbox{where}\quad K(x,x)=\<x\vert\frac{1}{ \Delta^2\!-\!
\mg^2\Delta}\vert x\>
\eqno(5.6)
$$
is the Green function on the space of functions which integrate
to zero. Using the eigenmodes (5.3) of $-\Delta$ and observing
that the excited modes span the space of functions $\phi$
permitted in the path integral (5.5a), we obtain
$$
K(x,y)=\sum_{m\neq 0}\frac{\phi^\dagger_m(x)\phi_m(y)}{ \mu_m^2+\mg^2\mu_m}
\Longrightarrow K(x,x)=\frac{1}{\mg^2 V}\sum_{m\neq 0}\big(\frac{1}{ \mu_m}-
\frac{1}{\mu_m+\mg^2}\big),
$$
where the eigenvalues $\mu_m$ of $-\Delta$ are the ones in (3.4) with
$a_0$ and $a_1$ set to zero. Using the identity
$$
\sum_{m_1=-\infty}^\infty \frac{e^{2\pi i m_1\Phi}}{ a^2+m_1^2}=
\frac{\pi}{ a}\frac{\cosh (\pi a[1-2\Phi])}{ \sinh (\pi a)}\quad(\Phi\geq 0)
\eqno(5.7)
$$
for $\Phi=0$ the summation over $m_1$ can be carried out and one
obtains ($n\!=\!m_1$)
$$
\mg^2 K(x,x)=\frac{1}{ \mg^2V}-\frac{\coth(L\mg/2)}{ 2\beta\mg}+\frac{\tau}{ 12}
+\frac{1}{ 2\pi}\sum_{n>0}\Big[\frac{\coth(n\pi \tau)}{ n}
-\frac{\coth(\pi\tau\xi(n))}{ \xi(n)}\Big]\eqno(5.8)
$$
where $\xi^2(n)=n^2+(\beta\mg/2\pi)^2$. The first sum on the right
hand side can be expressed in terms of Dedekinds function as
$$
\sum_{n>0}\frac{1}{ n}\big(\coth(\pi\tau n)-1\big)
=2\sum_{n,r>0}\frac{e^{-2\pi\tau nr}}{ n}=-2\sum_r\log\big(1-e^{-2\pi\tau r}
\big)=-2\big(\frac{\pi \tau}{ 12}+\log\eta(i\tau)\big)
$$
and thus cancels against the third term on the right hand side in (5.8)
and the $\vert\eta^2\vert$ in (5.6). Also note the $V$-dependent
exponential factor in (5.6) cancels against the first term on the right
hand side in (5.8) so that finally
$$
\<\bar\Psi P_\pm\Psi\>=-\frac{1}{\beta}\exp\Big[-\frac{\pi}{ \beta\mg}
\coth(\ha L\mg)\Big]\,e^{F(\beta\mg)}\,e^{-2H(\beta\mg,\tau)}\eqno(5.9a)
$$
where
\[
F(x)=\sum_{n>0}\big(\frac{1}{ n}-\frac{1}{ \sqrt{ n^2+(x/2\pi)^2}}\big)\hskip17mm\]
\vskip-1mm
\[H(x,\tau)=\sum_{n>0}\frac{1}{ \sqrt{n^2+(x/2\pi)^2}}\cdot
\frac{1}{ e^{\tau\sqrt{(2\pi n)^2+x^2}}-1}.
\eqno(5.9b)
\]
The formula (5.9) is the exact form of the two point functions
we have been aiming at.
This formula simplifies considerably if we let
$L\to \infty$, in which case $H(.,.)\to 0$ and $\coth(.)\to 1$ so that
$$
\<\bar\Psi \Psi\>=-\frac{2}{ \beta}\exp\Big[-\frac{\pi}{ \beta\mg}
\Big]\,e^{F(\beta\mg)},\eqno(5.10)
$$
where we have used that $P_+\!+\!P_-=Id$. \pan
The formula (5.9a) for the finite temperature
chiral condensate on a finite intervall and the corresponding
formula (5.10) after the limit $L\to\infty$ has been taken, are
the main results of this paper. Let us now derive the low - and
high temperature limits in turn:\pan
To study the low temperature limit of the chiral condensate we
regularize the indiviudal sums in (5.9b) as \cite{kirsten}
\[
\sum \frac{1}{n^{1+s}}\sim \frac{1}{ s}+\gamma+O(s)\hskip15mm\]\vskip-4mm
\[\sum \frac{1}{ (n^2+(x/2\pi)^2)^{\ha+s/2}} \sim
\frac{1}{s}-\frac{\pi}{ x}-\log\frac{x}{ 4\pi}+2\sum_{l=1}^\infty K_0(lx)+O(s)
\eqno(5.11)
\]
where $\gamma=0.57721...$ is Eulers constant. The difference is finite
as $s$ tends to zero and inserting it into (5.9) and (5.10) yields
$$
\<\bar\Psi \Psi\>=-\frac{\mg}{ 2\pi}e^\gamma\,e^{2I(\beta\mg)}\qw
I(x)=\int\limits_0^\infty \frac{1}{1-e^{x\cosh(t)}}dt,\eqno(5.12)
$$
where the last integral represents (up to a sign) just the last
sum in (5.11). This altenative representation of the chiral
condensate for $L=\infty$ has been derived in
the bosonised Schwinger model on the cylinder in \cite{manton}.
Since the integral $I$ tends to zero for low temperatures we find
$$
\<\bar\Psi \Psi\>\longrightarrow -\frac{\mg }{ 2\pi}\;e^{\gamma}\qquad\hbox{for}
\quad T\to 0,
\eqno(5.13)
$$
and this is the well known result for the zero temperature Schwinger
model \cite{nielsen}.\pan
To study the high temperature limit we observe
that $F(x)$ tends to zero as $x\ll 1$ or if the temperature is
big compared to the induced photon mass. In this region
$$
\<\bar\Psi \Psi\>\longrightarrow -2T\;e^{-\pi T/\mg}\qquad\hbox{for}
\quad T\to\infty.
\eqno(5.14)
$$
Thus the chiral condensate vanishes exponentially fast for
$T\gg \mg$.\pan
For the intermediate temperatures $0<T\sim \mg$ we must retreat
to numerical methods to evaluate the infinite sum which defines the
function $F$ appearing in (5.10). The result of the numerical calculations
are depicted in Figure 1.\pan
Note how the function $\<\bar\Psi\Psi\>(T)$ resembles the behaviour
of an order parameter in a system which suffers a second order
phase transition. However, the chiral condensate does not really
vanish at any finite temperature, it tends to zero 'only' exponentially,
contrary to the phase transition case. Thus in a strict sense the
chiral symmetry remains broken at all finite temperature.

\section{Further Correlation Functions}
In this section we determine the correlation functions
of Wilson loops, products of the field strength, thermal Wilson loops and gauge invariant fermionic two-point functions.
Clearly, for the purely bosonic operators only the zero-instanton sector
yields non-vanishing expectation values. Since
in this sector $A=$harmonics$+\eps\pa \phi$ one has
$$
\<O[A]\>=\sqrt{2\tau}\frac{\int \cd\phi\;e^{-\Gamma[\phi]}\int d^2h\,
\vert\Theta\vert^2\;O[\phi,h]}{ \int \cd\phi e^{-\Gamma[\phi]}},\eqno(6.1)
$$
for an operator $O[A]$. Here
we have used the second formula in (5.5c) and $\Theta$ is the
theta-function in (3.6). Hence, up to the $h$-integration we remain
with Gaussian integrals with respect to the effective action (5.5b).
\subsection{Correlation functions of the field strength}
The $n$-point functions of the field strength $E=\Delta\phi$ are
particularly simple, since $E$ does not depend on the harmonic pieces
in the gauge potential. Because of the symmetry $\phi\to -\phi$ of $\Gamma$
all expectation values of odd powers of $E$ vanish. The $2n$-point
functions can be calculated from
$$
\<E(x_1)\cdots E(x_{2n})\>=\frac{\int \cd \phi\, e^{-\Gamma[\phi]}
E(x_1)\cdots E(x_{2n})}{ \int \cd \phi\, e^{-\Gamma[\phi]}}
=\Delta_1\cdots \Delta_{2n}\<\phi(x_1)\cdots \phi_{2n})\>.
$$
and applying Wick's theorem to the last expectation value
allows one to express them in terms of the 2-point function
$$
\<E(x)E(y)\>=\Delta_x\Delta_y\<\phi(x)\phi(y)\>=\Delta_x\Delta_y
K(x\!-\!y,\mg),
$$
where the Green function $K$ corresponding to $\Gamma$ has been introduced
above (5.7). Since $-\Delta \phi_m=\mu_m\phi_m$ we easily find that
$$
\<\dots\>=\sum_{m\neq 0}\phi_m^\dagger(x)\phi_m(y)\big[1-\frac{\mg^2}{
\mu_m+\mg^2}\big]=-\Delta_x G(x-y,\mg).\eqno(6.2)
$$
Now we use the (almost) completeness of the excited eigenmodes
$\phi_n$ and obtain
$$
\<E(x)E(y)\>=\delta(x-y)-\frac{1}{V}-\mg^2 G(x\!-\!y;\mg)\eqno(6.3)
$$
where $G$ denotes the massive Klein-Gordon propagator. This shows
that $E$ is (up to contact terms) just a free massive pseudo-scalar field.
Using (5.7) the propagator can be written as
$$
G(\xi;\mg)=\frac{1}{4\pi}\sum_{n=-\infty}^\infty
\frac{\cosh\big[\pi\tau\xi(n)(1-2\xi^1/L)\big]\,e^{2\pi i n\xi^0/\beta}}{
\xi(n)\sinh\big[\pi\tau \xi(n)\big]}-\frac{1}{ \mg^2 V},
\eqno(6.4a)
$$
where $\xi(n)$ has been defined below (5.8). In the limit
where $L$ tends to infinity this simplifies to
$$
G(\xi;\mg)=\frac{1}{4\pi}\sum_n
\frac{e^{-2\pi[\xi(n)\xi^1-in\xi^0]/\beta}}{ \xi(n)}.\eqno(6.5)
$$
This shows explicitely that the $2$-point function of the field strength falls off exponentially.\pan
\subsection{Wilson loops}
Since loop integrals of local Wilson loops (that is loops without windings)
also do not depend on the harmonic pieces one finds
$$
\<e^{ie\oint A}\>=\frac{\int\cd\phi\, e^{-\Gamma[\phi]}\;\exp\Big(ie\int_\cd
\Delta\phi\Big)
}{ \int\cd\phi \,e^{-\Gamma[\phi]}},
$$
where the loop encloses the region $\cd$ and we have used that
the loop integral is equal to the flux of $E=\Delta\phi$ through $\cd$.
Using (6.3) the resulting Gaussian integral yields
\[
\<e^{ie\oint A}\>=\exp\Big(-\frac{e^2}{2}\int_{\cd\times\cd}
\<E(x)E(y)\>d^2x d^2y\Big)\hskip16mm\]
\[ \hskip18mm=
\exp\Big(-\frac{e^2}{2}\Big[A_\cd-\frac{A_\cd^2}{ V}-\mg^2\int_{\cd\times\cd}
G(x\!-\!y,\mg)\Big]\Big),\eqno(6.6a)
\]
where $A_\cd$ is the area enclosed by the loop. For example, for
a rectangle $\cd$ one finds for $L=\infty$
$$
\<e^{ie\oint A}\>=
\exp\Big(-\frac{e^2 A_\cd}{ 2}\big[1-\mg^2\int_{2\cd}G(\xi;\mg)d^2\xi\big]\Big).
\eqno(6.6b)
$$
For low temperatures (compared to the induced photon mass) and loops with
edges long compared to the inverse photon mass we may use the
zero-temperature propagator $G(\xi,\mg)=K_0(\mg \vert\xi\vert)/2\pi$ in
(6.6b) for which the integral in the exponent is, up to an exponentially small term, just $\mg^{-2}$. It
follows then that at low temperature the interaction between two charges is exponentially small\pan
\subsection{Thermal Wilson loops}
The expectation values of a string of thermal Wilson loops
$$
\cp (u)=\exp\Big(ie\int\limits_0^\beta\,A_0(x^0,u)dx^0\Big)\eqno(6.7)
$$
depend on the harmonics because a thermal Wilson loop has winding number one.
Since $\int A_0 dt=2\pi h_0-\pa_1\!\int\phi dx^0$ the $h$-integral
for a string of $q$ loops becomes
$$
\int d^2h\vert\theta\vert^2\,e^{2\pi iqh_0}=\sqrt{\frac{1}{ 2\tau}}
e^{-\pi q^2/2\tau}.
$$
The remaining $\phi$-integration yields
$$
\<\prod_{i=1}^q\cp (u_i)\>=e^{-\pi q^2/2\tau}\exp\Big[
\frac{e^2}{2}\Big(\int d^2x d^2y\sum_{i=1}^q\delta(x^1\!-\!u_i)\;\pa^2_{x^1}
K(x,y)\sum_{i=1}^q\delta(y^1\!-\!u_i)\Big)\Big].
$$
Inserting the explicit form of $K$ (see above (5.7)) one finds the
following one- and two-point functions
\[
\<\cp (0)\>=\exp\Big(-\frac{\pi\beta\mg}{4}\coth(\ha L\mg)\Big)\]
\[\<\cp (u)\cp(0)\>=\<\cp(0)\>^2\exp\big(-\ha\pi\beta\mg S(u)\big)\hskip8mm
 \eqno(6.8a)
\]
where $S$ denotes the function
$$
S(u)=\frac{\cosh\ha \mg (L-2\vert u\vert)}{ \sinh(\ha\mg L)}.\eqno(6.8b)
$$
The normalized higher n-point functions
$$
\cp (u_1,\dots,u_q)\equiv \frac{\<\cp (u_1)\cdots \cp(u_q)\>
}{ \<\cp (0)\>^q}\eqno(6.9)
$$
are then just products of the normalized two-point function
$$
\cp (u_1,\dots,u_q)=\prod_{i<j}\cp (u_i,u_j).\eqno(6.10)
$$
Since $-T\log\cp(u_1,\dots,u_q)$ is to be intepreted as the (zero energy
subtracted) free energy of $q$ static charges at positions $u_1,\dots,u_q$,
(6.10) shows that the free energy of $q$ charges is just the sum
of the free energies of the individual pairs.\pan
For $L=\infty$ the one and two-point functions (and thus the $q$-point
functions) simplify to
$$
\<P(u)\>=\exp\big(-\pi\beta\mg/4\big)\qa
\cp(u,0)=\exp\big(-\ha\pi\beta\mg\,e^{-\mg\vert u\vert}\big),
$$
in complete agreement with the analytically continued one and two-point
functions obtained via bosonization in \cite{manton}.
Note in particular that the free energy of a single charge is finite
$$
\Delta F=-T\log\<\cp (0)\>=\frac{\pi\mg}{4}\eqno(6.11)
$$
and that the (zero energy subtracted) free energy of two
static charges falls off exponentially leading to a Yukawa force
$$
F=\frac{e^2}{2}\,e^{-\mg\vert u\vert}\eqno(6.12)
$$
between them. We see that the classical Coulomb force between
external charges is shielded.
\subsection{Gauge invariant fermionic two-point functions}

The gauge invariant chiral two-point functions
$$
S_\pm(x,y) =\<\Pb (x)\;e^{ie\int^x_y A_{\mu}d\xi^{\mu}}P_\pm
\Psi (y)\>\eqno(6.13a)
$$
may lead, via the LSZ reduction procedure, to chirality violating
scattering amplitudes. They can be calculated in the same way as the
chiral condensates (5.5a) which they must reproduce for coinciding points.
One obtains
$$
S_+(x,y) =\frac{-1}{ \beta}e^{-2\pi^2/e^2V}\;
\frac{\int d^2h \,\t\psi^\dagger_0(x)\t\psi_0(y)\;\int\cd \phi \;
e^{-\Gamma[\phi]-e(\phi(x)+\phi (y))+ie\int^x_y A}
}{ \int d^2h\vert \frac{1}{ \eta}\Theta\vert^2\int \cd\phi \,e^{
-\Gamma[\phi]}},\eqno(6.13b)
$$
where the zero mode $\t\psi_0(x)$ has been computed in (4.4).
The integration over the harmonics yields
$$
\int d^2h\,\t\psi^\dagger_0(x)\t\psi_0(y)\;e^{ie\int^x_y A} =
\frac{C}{\sqrt{2\tau}} \ e^{-\frac{\pi}{{2V}}(x-y)^2}\;e^{-i\int^x_y
\eps_{\mu\nu}\pa_\nu \phi d\xi^{\mu}},\eqno(6.14)
$$
where $C$ is some normalisation constant. The $\phi$-integration is again
Gaussian but due to the phase factor in (6.14)
involves derivatives of the bosonic Green function:
\[
\int{\cal D\phi}\;e^{-\Gamma[\phi]-e(\phi(x)+\phi (y))-ie\int\eps_{\mu\nu}
\pa_\nu \phi d\xi^{\mu}} =\tilde C \exp\bigl[K(0,0)+K(x,y)\bigr]\times\hskip20mm\]\vskip-9mm
\[\hskip72mm\exp\big[-\ha\int\limits^y_x d\xi^\mu d\eta^\alpha\eps_\mu^\nu\eps_\alpha^\beta
\pa_{\xi^\nu}\pa_{\eta^\beta}K(\xi,\eta)\big],
\eqno(6.15)
\]
where 
$\tilde C$ is again some constant. For simplicty we shall first consider the zero
temperature and
infinite volume limit only. There $K$ depends only on the distance of its
two arguments and the integral on the r.h.s of (6.15) vanishes identically.
Hence we are then left with the computation of the Green function $K(x,y)$.
This Green function,
which is the difference of the masseless- and the massive Klein-Gordon
propagator (see 5.7) simplifies to
$$
K(x,y) = \frac{1}{{\mg^2}}\bigl[G(r;\mg=0)-G(r;\mg)\bigr]=\frac{{-1}}{{2\pi\mg^2}}
\bigl[\log (\mu r)+K_0(\mg r)\bigr],\eqno(6.16)
$$
where $r=\vert x-y\vert$ and $\mu$ is an infrared regularisation which
would be undefined on the Euklidean space.
Here it is fixed by demanding that $S_\pm(x,x)$ reproduces
the chiral condensate (5.13). Finally substituting (6.14 -16) into (6.13)
yields
$$
\<\Pb (x)\;e^{ie\int A}P_+\Psi (y)\> =
-\sqrt{\frac{{m_\ga e^{\ga}} }{{2}}}\frac{ \exp[-\ha K_0(\mg \vert x-y\vert)]}{{2\pi \sqrt{\vert x-y\vert}}}.\eqno(6.17a)
$$
For large separations $K_0(\mg r)$ vanishes exponentially so that the
chiral two-point function is long range:
$$\<...\> \rightarrow -\sqrt{\frac{m_\ga e^{\ga}}{ 2}}\frac{1
}{{2\pi \sqrt{\vert x-y\vert}}},\eqno(6.17b)$$
for $\vert x-y\vert$ approaching infinity. Note that contrary to the
bosonic correlators which fall
off exponentially due to the mass gap in the spectrum, the chiral fermionic
$2$-point function falls of like $1/\sqrt{r}$. This is
due to the zero mode fermionic states in the Hilbertspace.\par
In a next step we will increase the complications by considering the case of again infinite volume
but finite temperature. This time however we must assume the difference of the two arguments of the
two-point function to be strictly spacelike (eg: $x^0-y^0 =0$) for then the integral on the r.h.s. of (6.15) vanishes again. This becomes clear if we notice that now only "timelike" derivatives of the bosonic Green function appear and these are zero for $x^0-y^0 =0$ as can be seen using the sum representation of $K(x,y)$ (above (5.6)). Hence we have again reduced the problem to the computation
of $K(x,y)$. However, as in the case of coinciding arguments (chiral condensate) we couldn't find
simple expressions representing the infinite sums for finite temperature. Therefore we again retreat
to numerical methodes to evaluate (6.15). The results are despicted in Figure 2. We will not reproduce
here the lengthy formulas but emphasize that for non-zero temperature and for $|x^1-y^1|\rightarrow\infty$ , $S_+(x,y)$ falls of like $\exp [-2\pi T|x^1-y^1|]$. This is the expected behaviour at large temperature.

\section{Discussion and Conclusions}
Returning to (5.6) and noting that
$$
\frac{\vert \eta(i\tau)\vert^2}{ \beta}=\frac{\eta(i\tau)\eta(i/\tau)}{ \sqrt{V}}
\qquad (\tau=L/\beta)\eqno(7.1)
$$
is left invariant under an exchange of $\beta$ with $L$, and that
the same is true for $K(x,x)$, we may view (5.10) as dependence of the
zero-temperature chiral condensate on the spatial extension $L$. Thus
Figure 1 can be intepreted as the change of $\<\bar\Psi\Psi\>(L)$
due to finite (spatial) size effects. In particular, if the compactified
spatial size shrinks to zero, the chiral condensate behaves as
$$
\<\bar\Psi\Psi\>(L)\longrightarrow -\frac{2}{ L}\,e^{-\pi/L\mg}
\qquad\hbox{for}\quad L\to 0\eqno(7.2)
$$
and vanishes exponentially (and non-analytically).\par
In the decomposition (5.4) of the bosonic path integral we have
integrated over different $U(1)$-principal bundles labelled
by the index $k$. We have assumed that the relative
normalization between functional integrals with different $k$
is one. To recover the $\theta$-vacuum structure \cite{loewenstein} we must allow
for a relative phase $\exp(i\theta)$ between
the sectors $k$ and $k\!+\!1$ (similarly as in the path integral for
a particle on a punctured plane). Then the expectation values
$\<\bar\Psi P_\pm\Psi\>$ would pick up the phases $\exp(\pm i\theta)$
and thus the right hand sides of (5.10) and (5.12-14) would be multiplied
by $\cos(\theta)$. The same happens if one adds the explicit $\theta$-dependent
term $i\theta\Phi/2\pi$ to the classical action (1.1). \par
In the conventional path integral solution of the Schwinger model
on the plane the zero-temperature result (5.13) is derived indirectly,
via clustering of the $4$-point function (1.17). This seems to
be necessary since a (naive) direct calculation of the chiral
condensate yields the wrong result $\<\bar\Psi\Psi\>=0$. This may
not come as a surprise since one integrates only over potentials
with vanishing fluxes and on the other hand we have seen that on the
torus only configurations with fluxes $\pm 2\pi$ contribute to the
expectation value (5.13).
Integrating over all gauge potentials should yield the correct result
(5.13). More precisely, on the plane a gauge potential can be decomposed as
$$
A_\mu=\tilde A_\mu-\eps_{\mu\nu}\pa_\nu\phi,\eqno(7.3a)
$$
where $\tilde A$ is a vortex field
$$
\tilde A_\mu=-\frac{1}{ 2\pi}\frac{\Phi}{a^2+r^2}\eps_{\mu\nu}x^\nu,\qquad
\Phi=\int E\eqno(7.3b)
$$
and $\delta A_\mu=-\eps_{\mu\nu}\pa_\nu\phi$ a local fluctuation
with vanishing flux about the vortex potential.
One can show that $\tilde\di$ possesses exactly $\vk$ zero modes,
where $\vk$ is the integer part of $\vert\Phi\vert/2\pi$ \cite{musto}
(besides these zero modes there is also a resonance state).
Since only potentials with $\vk =1$ contribute to the chiral condensate
one obtains
$$
\<\bar\Psi P_+\Psi\>=
\frac{\int\limits_{1<\Phi\leq 2}d\Phi \;\rho(\Phi)
\int\cd\phi\, e^{-\Gamma[\phi]}\tr\big(\psi^\dagger P_+\psi\big)}{
\int\limits_{-1\leq\Phi\leq 1}d\Phi \;\rho(\Phi)
\int\cd\phi\, e^{-\Gamma[\phi]}},\eqno(7.4)
$$
where we have allowed for a Jacobian
$\rho$ of the transformation (5.2) (without harmonics $h$). Again
the effective action $\Gamma$ contains two pieces, namely
a local part coming from the chiral anomaly and
a global vortex determinant. Since on the plane there aren't
different topological sectors it would be interesting to
see how the $\theta$-parameter emerges in such a calculation.

\vskip4mm
\textbf{Acknowledgments:}
The authors wish to thank G. Felder and C. Schmid for helpful discussions. In addition A. Wipf thanks J. Balog, M.P. Frey, K. Kirsten and E. Seiler for useful discussions.

\end{document}